%% file: main.tex
\documentclass[sigconf, nonacm]{acmart}
\usepackage{subcaption}
\usepackage{multirow}
\usepackage{enumitem}

\begin{document}

\title{Parallelization Strategies for Dense LLM Deployment: Navigating Through Application-Specific Tradeoffs and Bottlenecks}

\author{Burak Topcu}
\email{bvt5283@psu.edu}
\affiliation{%
  \institution{The Pennsylvania State University}
  \city{State College}
  \state{Pennsylvania}
  \country{USA}
}

\author{Musa Oguzhan Cim}
\email{mtc5693@psu.edu}
\affiliation{
  \institution{The Pennsylvania State University}
  \city{State College}
  \state{Pennsylvania}
  \country{USA}
}
\author{Poovaiah Palangappa}
\email{Poovaiah.Palangappa@amd.com}
\affiliation{
  \institution{Advanced Micro Devices}  
  \city{San Jose}  
  \state{California}  
  \country{USA}
}
\author{Meena Arunachalam}
\email{Meena.Arunachalam@amd.com}
\affiliation{
  \institution{Advanced Micro Devices}  
  \city{Portland}  
  \state{Oregon}  
  \country{USA}
}
\author{Mahmut Taylan Kandemir}
\email{mtk2@psu.edu}
\affiliation{
  \institution{The Pennsylvania State University}  
  \city{State College}  
  \state{Pennsylvania}  
  \country{USA}
}

\begin{abstract}
Breakthroughs in the generative AI domain have fueled an explosion of large language model (LLM)–powered applications, whose workloads fundamentally consist of sequences of inferences through transformer architectures. Within this rapidly expanding ecosystem, dense LLMs—those that activate all model parameters for each token generation—form the foundation for advanced expert-based variants. Dense models continue to dominate because of their strong generalization ability, scalability, ease of fine-tuning, and versatility across diverse tasks. In LLM inference systems, performance is mainly characterized by latency, response time, and throughput (i.e., tokens generated per unit of time). Latency and throughput are inherently coupled: optimizing for one often comes at the expense of the other. Moreover, batching strategies and parallelism configurations, which are essential when dense model parameters exceed device memory capacity, can significantly affect both latency and overall system throughput. This paper (i) investigates the workloads of two representative dense LLMs—Llama-3.1-70B and Llama-3.1-405B, focusing in particular on intra-node parallelization schemes, (ii) analyzes how input characteristics, batching, and parallelism strategies influence latency flexibility and the latency–throughput tradeoff, and (iii) identifies key performance bottlenecks that inform design choices for meeting service-level agreements (SLAs) and sustaining inference quality. Our empirical evaluations reveal that Tensor Parallelism (TP) improves the latency objectives while Pipeline Parallelism (PP) is better-suited for throughput-oriented applications. We highlight that their hybrid usage by controlling the TP and PP degrees provides control over the latency-throughput interplay.
\end{abstract}


\maketitle

\input{sections/introduction}

\input{sections/background}

\input{sections/sim_validation}
\input{sections/methods}

\input{sections/evaluations}

\input{sections/disc_and_conc}

\bibliographystyle{ACM-Reference-Format}
\bibliography{sample-base}

\end{document}

%% file: sections/introduction.tex
\section{Introduction} \label{sec:introduction}

The rapid proliferation of generative AI, particularly through large language models (LLMs), has profoundly transformed various aspects of modern life, streamlining a wide range of real-world tasks such as code generation~\cite{streamGen, codeGen}, document summarization~\cite{docSummarization}, conversational agents~\cite{chatbots}, and even clinical diagnosis~\cite{med_diag_app1, apps_eval3}. Empirical evidence increasingly suggests that larger models, especially those with extended context windows, are better equipped to capture the complex, hierarchical, and highly nonlinear structures inherent in natural language~\cite{fewShotLearn_scaling1, emergAbil_scaling2, scalingLaws_3, hoffman_scaling4}. Although recent advances in Mixture-of-Experts (MoE) and other sparse architectures have improved computation efficiency by selectively activating subsets of parameters, dense models, where all parameters contribute to every token generation, remain indispensable. They not only form the foundation for large-scale pretraining and fine-tuning, but also serve as the basis for downstream transformations such as pruning~\cite{dense2sparse}, quantization~\cite{dense2quant}, and expert development~\cite{dense2MoE,dense2MoE_2}. Moreover, their inherent scalability, robustness, and versatility continue to make dense architectures the cornerstone of modern generative AI systems.

Popular state-of-the-art dense LLMs, such as variants of GPT-4~\cite{GPT4}, GPT-3~\cite{GPT3}, Llama3.1~\cite{llama3}, Mistral Large 2~\cite{mistrallarge2}, and some of Qwen-2.5~\cite{qwen2_5} models, have hundreds of billions of model parameters and support context lengths up to tens of thousands of tokens. However, generating tokens for a batch of requests while serving inferences for large contexts through large models demands considerable memory to store model weights as well as the runtime key-value (KV) cache, which grows dynamically along with autoregressive decode phases, and often results in exceeding the memory capacity of a single GPU device. For example, a system serving inferences through the Llama3.1-405B model~\cite{llama3} in FP8 quantization necessitates \textasciitilde405GB of memory space to only accommodate model weights, and a single GPU--even cutting-edge server-class models, such as MI325x~\cite{mi325x} or H200~\cite{h200}--cannot serve this model at 8-bit or higher quantization due to physical memory insufficiency. Thus, deploying such models for inference purposes, which typically mandates cutting-edge compute infrastructure generally equipped with multiple GPUs or TPUs and a scalable interconnect between them, can often lead to LLM inference serving setups becoming markedly costly.

To tackle the resource-intensive nature of LLM inference services, recent studies have proposed offloading model parameters and KV caches to host memory when GPU memory capacity is exceeded, retrieving them back on-demand~\cite{flexGen, deepspeedInf}. Despite the reduced cost of inference serving, this approach often falls short of achieving high throughput and low latency due to the communication overheads introduced between the host and accelerators. To further alleviate memory pressure, quantization, compression, and pruning have emerged as promising alternatives that allow large models and their growing KV cache footprints to fit within more constrained memory spaces. Quantization~\cite{quant_flexgen, quant_int8} reduces the numerical precision of the model parameters and KV cache entries (e.g., FP6 and FP4~\cite{mxfp4}), allowing one or fewer accelerators to serve inferences with minimal loss in accuracy. On the other hand, pruning~\cite{prune_ThinK, prune_DynContPr, prune_KVPruner} aims to identify a region of model parameters or the KV cache, such as a subset of transformer blocks or a partial segment of KV cache entries, with minimal impact on the inference output and removes them with tolerable accuracy tradeoffs. Additionally, compression~\cite{compres_gear, compress_AdapComp, compress_squeezeLLM} relaxes the memory pressure by uniformly or selectively compressing and decompressing the model or KV cache during inference, albeit at the cost of additional compute overhead. Although these techniques effectively mitigate resource constraints associated with serving large models, none fully satisfy the stringent latency and throughput requirements of modern LLM-serving workloads without compromising Quality-of-Service (QoS).

Parallelization techniques, in contrast, alleviate memory pressure from model weights and KV cache by distributing them across multiple GPUs, creating substantial opportunities to improve latency and throughput without compromising QoS. Common model parallelization strategies, such as Tensor Parallelism (TP)~\cite{megatronLM} and Pipeline Parallelism (PP)~\cite{pipeDream}, exhibit diverse tradeoffs as they use distinct methods to parallelize LLM inferences. For example, both TP and PP, which shard the internal transformer layers and partition the transformer blocks across multiple devices, respectively, enable large model deployment, constrained by single-device memory limitations, and relax the per-device memory requirement, presenting the operation with large batches. We believe that parallelism techniques for LLM inference systems emerge as the most promising way to achieve high throughput or latency sensitivity for LLM-powered applications, as well as ensuring the inference quality. Still, the interplay between parallelization schemes, distinct resource usages resulting from parallelizing transformer workloads differently, model and diverse input characteristics, and batching preferences, as well as their collective influence on latency and throughput, remains underexplored, hindering the achievement of performant inference services for LLM applications.

To the best of our knowledge, no prior work has systematically evaluated the performance of various parallelism strategies under different system and application constraints, nor identified the key bottlenecks that emerge across these strategies. Motivated by this, the primary objective of this study is to thoroughly analyze, through an in-house simulator, the opportunities and challenges associated with data~\cite{dp2_nnt} and model parallelism techniques, such as tensor~\cite{megatronLM} and pipeline~\cite{pipeDream} parallelism, along with their hybrid compositions. In this study, we (i) characterize how transformer workloads map onto GPU-equipped node systems under different parallelization schemes and depths; (ii) quantify latency and throughput trends across a range of parallelization configurations, batching strategies, and input characteristics when serving inference with the Llama 3.1–70B and 3.1–405B models; and (iii) identify the optimal parallelization setups that deliver stringent latency and high throughput, while diagnosing the key limitations in both inference workload behavior and system capabilities that constrain further improvements. We finally (iv) discuss the broader applicability of our findings and insights to multi-node systems, sparse and expert-based models, and how more advanced forms of parallelization, such as expert parallelism~\cite{expertPar}, can impact the latency and throughput objectives for dense models in such settings. The main {\bf contributions} of this work can be itemized as follows: 

$\bullet$ We demonstrate the kernel-level breakdown of a typical dense LLM transformer workload and the mapping of transformer layers under various parallelization schemes, their hybrid variants, and scaling depths onto GPUs within a node environment. 

$\bullet$ We quantify the latency and throughput changes driven by batching and parallelization preferences, together with distinct model and input characteristics, using both short- and long-context representative datasets, all possible batch sizes that fit into the system, and diverse parallelization strategies, encompassing hybrid variants, with a range of their respective degrees.

$\bullet$ We analyze the latency flexibility and latency-throughput tradeoff, satisfying a wide spectrum of application objectives, and identify limitations that prevent further improvements of the application performance objectives. 

The remainder of this paper is organized as follows: Section~\ref{sec:background} outlines the LLM inference workflow, introduces the target evaluation models, and reviews the current literature, motivating a detailed evaluation of parallelism techniques. Section~\ref{sec:simval} validates the execution correlation between silicon and the in-house simulator, and Section~\ref{sec:parStrategies} details the parallelization strategies and explores their workload mapping in a multi-GPU system. Section~\ref{sec:evaluations} presents a comprehensive evaluation of latency and throughput under various configurations of parallelization, batching, model, and input datasets. Lastly, Section~\ref{sec:disc} discusses the applicability of our analysis to both multi-node systems and expert-based models, and Section~\ref{sec:conc} concludes with key insights and promising trends revealed by our findings.

%% file: sections/background.tex
\section{Background and Related Work} \label{sec:background} 
\subsection{Large Language Models} 
Large language models (LLMs) are deep neural networks pretrained on massive text corpora and built as a sequence of {\it transformer blocks}. Each transformer block is the core computational unit: it applies attention and feed-forward operations that capture contextual relationships among tokens. When an inference request is issued, the model processes it in two phases: (i) a {\bf prefill phase}, where all input tokens are processed in parallel, and (ii) a {\bf decode phase}, where new tokens are generated autoregressively, one at a time, until an end-of-sequence (EoS) token is reached. In both the phases, every request in the batch is propagated through the entire stack of transformer blocks in order, because each block consumes the output of the previous one.

A transformer block mainly consists of QKV projection, rotary positional encoding (RoPE), attention, output projection, residual addition \& normalization layers, and a feedforward network (FFN), which might consist of multiple FFN sub-networks (i.e., experts). QKV projection layer transforms the input embeddings into Query (Q), Key (K), and Value (V) vector components prior to the attention layer. The \textit{attention} layer~\cite{attention} measures the attention scores and converts them into a probability distribution, denoting the pairwise relevance between tokens. Following that, an output projection layer maps the attention output back to the hidden dimension. FFN block--typically a single-layer or a two-layer gated feed-forward network incorporating a nonlinear activation function between linear projections~\cite{glu_variants} (expert-based models have multiple FFNs~\cite{expertPar})--learns complex contextual patterns through fully connected deep neural networks. Lastly, the residual addition layer captures the gradient changes following either the attention or FFN layer, and the normalization layer stabilizes the output before passing it to either FFN or the next transformer block. 

\subsection{Overview of Llama 3.1-70B/-405B Models}
Llama 3.1~\cite{llama3} represents the current state of the art in open-weight LLMs, offering performance comparable to leading proprietary models while remaining fully accessible for reproducible experimentation and systems research. It actually comprises a family of pretrained and fine-tuned models with parameter sizes of 8B, 70B, and 405B, achieving QoS comparable to current leading models such as GPT-4, GPT-3.5, and Mixtral-8×22B Turbo on several benchmark tasks~\cite{llama3_comp}. Llama 3.1 models support a context length of 128K, allowing comprehension of significantly longer inputs and demonstrating strong capabilities in multilingual understanding, reasoning, mathematics, coding, and summarization. Table~\ref{tab:model_specs} summarizes the specifications of 70B and 405B versions, determining the key inference workload characteristics influencing latency and throughput, in addition to varying input conditions and system configurations. The 70B version features smaller hidden and intermediate dimensions (i.e., the up-projection in the FFN layer), as well as fewer attention heads and transformer blocks, compared to the 405B variant.

\begin{table}[H]
  \centering
  \vspace{-1mm}
  \caption{Architecture specifications for Llama 3.1-70B/-405B.\label{tab:model_specs}}
  \vspace{-3mm}
  \begin{tabular}{l c c}
    \toprule
                        & {\textbf{70B Model}} & {\textbf{405B Model}} \\
    \midrule
    Number of transformer blocks    & 80     & 126    \\
    Hidden dimension size           & 8192   & 16384  \\ 
    Intermediate dimension size     & 28672  & 53248  \\
    Number of attention heads       & 64     & 128    \\ 
    Attention head dimension        & 128    & 128    \\
    Total number of parameters      & 70B    & 405B   \\
    \bottomrule
  \end{tabular}
  \vspace{-2mm}
\end{table}

\subsection{Related Works} 
{\bf Design Knobs Controlling Latency \& Throughput:~} Most inference system designs comprise a host and a set of accelerators~\cite{orca,flexGen}. The host manages communication with external components and runs LLM-serving software infrastructure, which governs the metadata and control flow of inference execution, and dispatches kernels to accelerators, whereas the interconnected accelerators (typically GPUs) perform the core inference computations. The primary challenges in achieving application objectives stem from the autoregressive nature of token generation and the large memory footprint of model parameters, KV caches, and intermediate activations. Therefore, efficient memory management, batching strategies, scheduling policies, and parallelization schemes constitute essential design aspects of inference systems, influencing service latency and system throughput. 

To reduce the memory demands of LLMs, several approaches have been proposed, such as model compression~\cite{compres_gear, compress_AdapComp, compress_squeezeLLM}, quantization~\cite{quant_flexgen, quant_int8}, and pruning~\cite{prune_ThinK, prune_DynContPr, prune_KVPruner}, alleviating the footprint of model parameters and KV caches. In addition, PagedAttention~\cite{vLLM} addresses memory fragmentation using a paging mechanism inspired by operating systems, and vAttention~\cite{vAttention} further minimizes the runtime memory allocation overhead of the KV cache, which grows at inference time, by leveraging the contiguity of logical memory. Furthermore, ORCA~\cite{orca} presents iterative and selective batching, which extends continuous batching with a more sophisticated design to initiate the inference of new requests upon their arrival. In comparison, dynamic batching~\cite{dynBatch_1, dynBatch_2} offers a more granular way to control resource utilization and SLA with memory-sensitive and application-specific batching opportunities. Moreover, scheduling policies such as ranking-based scheduling, which predicts the relative ranks of output lengths~\cite{rankBasedSch}, and speculative decoding~\cite{speculativeDec}, which estimates multiple output tokens in parallel, provide mechanisms to regulate SLAs and improve overall system throughput. In addition, DistServe~\cite{distServe} redesigns the system-level scheduling of inference workloads,  separating the prefill and decode phases across different accelerators (i.e., disaggregated prefill) to eliminate prefill–decode interference. ChunkedPrefill~\cite{chunkedPrefill} focuses on balancing the resource utilization of the prefill and decode phases, and in this regard, it proposes dynamically interleaving long prefills into multiple chunks and scheduling decodes between the interleaved prefill chunks to improve latency and throughput objectives. Current schedulers, such as POD-Attention and TensorRT-LLM~\cite{PODAttn,TensorRT_Chunk}, even form batches that include both chunked prefills and decode phases together to obtain optimal compute and memory co-utilization.

{\bf Parallelization Approaches for Large Model Deployment and Performance Control:} {\em Data Parallelism} (DP)~\cite{dp1_cnn, dp2_nnt} locates the replica of the model parameters on each inference serving instance, not incurring any inter-accelerator communication overhead, but consuming accelerator memory with model weights and limiting larger batches. DP provides scalability and is effective in optimizing the system throughput for applications without stringent latency constraints, when the model fits into GPU memory and sufficient space remains to accommodate large batches. {\em Tensor Parallelism} (TP)~\cite{megatronLM} shards within-transformer layers across accelerators such that accelerators process the same layer in parallel and aggregate the partial matrices within the transformer through \textit{all-reduce} operations. Despite incurring the overhead of all-reduce computations, TP can improve latency by dedicating more compute units to both the prefill and decode phases, as well as providing a larger memory space for both models that do not fit into a single accelerator and larger batches of requests. In comparison, {\em Pipeline Parallelism} (PP)~\cite{pipeDream, gpipe} partitions the transformer blocks across accelerators in proportion to the depth of parallelization, with each accelerator corresponding to a pipeline stage and running a set of transformers. Unlike TP, which concurrently processes the same batch on multiple accelerators, PP enables concurrent processing of multiple batches across distinct pipeline stages at inference time, thereby enhancing system throughput at inference time, rather than providing latency benefits. 

\begin{figure*}[!htb]
    \centering
    \begin{subfigure}{0.5\textwidth}
        \centering
        \includegraphics[width=\textwidth]{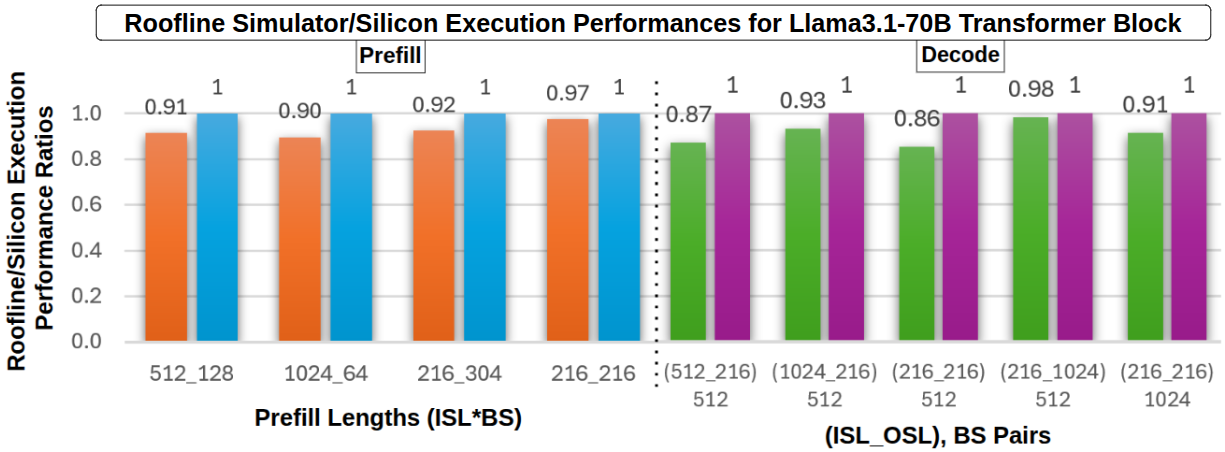}
        \caption{Normalized simulator latencies for prefill and decode relative to silicon for Llama 3.1-70B model.}
        \label{fig:llama3_70b_main}
    \end{subfigure}
    \begin{subfigure}{0.46\textwidth}
        \centering
        \includegraphics[width=\textwidth]{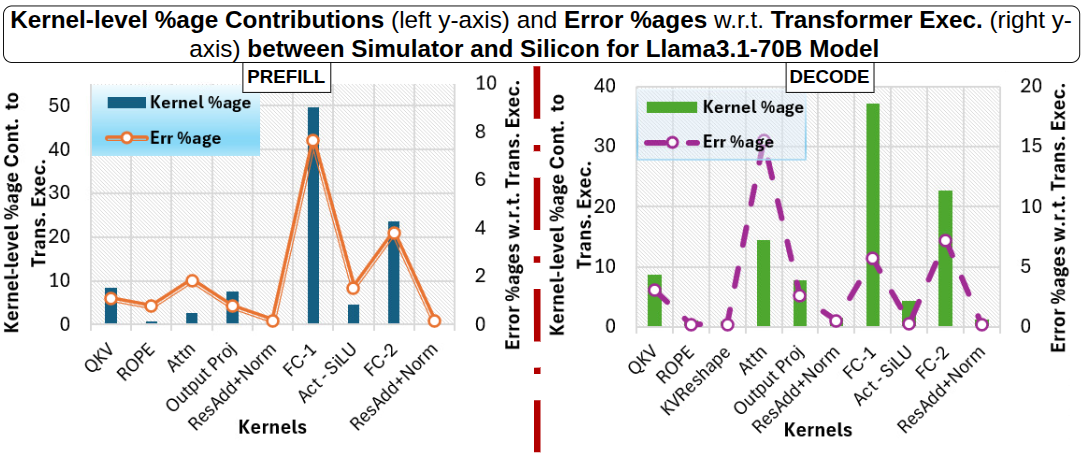}
        \caption{Contributions of individual kernels to the transformer execution, and kernel error percentages w.r.t. transformer duration.}
        \label{fig:llama3_70b_break}
    \end{subfigure}
    \hfill
    \begin{subfigure}{0.5\textwidth}
        \centering
        \includegraphics[width=\textwidth]{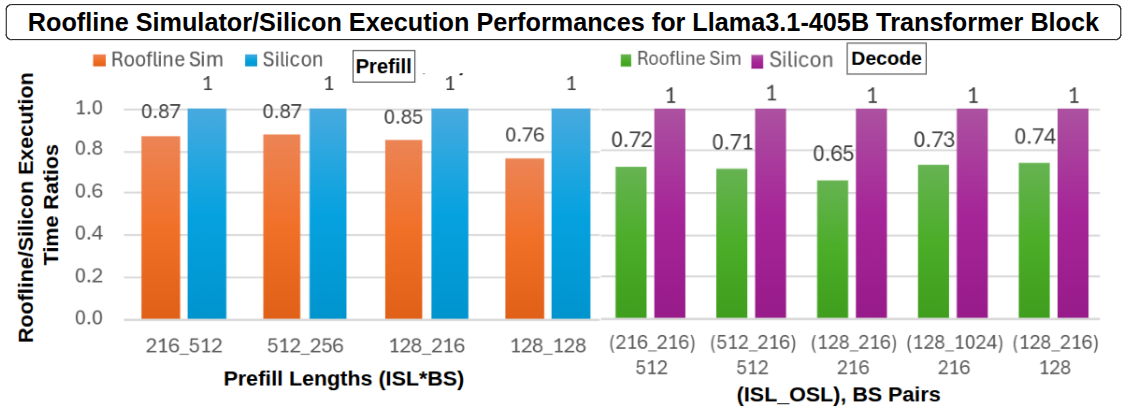}
        \caption{Normalized simulator latencies for prefill and decode relative to silicon for Llama 3.1-405B model.}
        \label{fig:llama3_405b_main}
    \end{subfigure}
    \begin{subfigure}{0.46\textwidth}
        \centering
        \includegraphics[width=\textwidth]{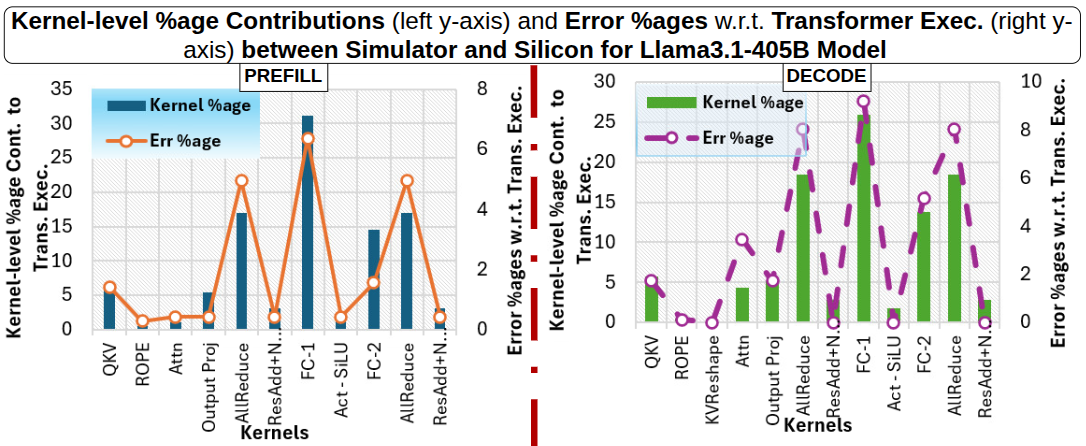}
        \caption{Contributions of individual kernels to the transformer execution, and kernel error percentages w.r.t. transformer duration.}
        \label{fig:llama3_405b_break}
    \end{subfigure}
    \vspace{-1mm}
    \caption{Figures (a,c) present simulator transformer pass times in \textit{prefill} and \textit{decode} normalized to silicon for \textit{Llama 3.1-70B/-405B} models. Figures (b) and (d) break down kernel execution times within transformer blocks in prefill and decode for the silicon, and reveal kernel-wise execution comparison errors w.r.t. transformer execution times on the right side y-axis.}
    \Description{Simulator validation.}
    \label{fig:llama_prefill_validation}
    \vspace{-3mm}
\end{figure*}

Beyond regular model-parallel approaches such as TP and PP, other parallelization schemes, in particular context~\cite{contextPar,parContextWind}, expert~\cite{gshard,effInferenceMoEs,DeepSpeed_MoE}, or hybrid use of multiple parallelism techniques~\cite{hybrid_training,alpa}, have been proposed by the prior art to address the memory bottleneck and tackle SLAs for LLM-powered applications. In parallel, previous research, such as~\cite{EvalMGPUInter,amd_infi_fabric,gpuTogpuComm2}, has evaluated modest GPU interconnects and revealed that topology preference, chiplet design, algorithm selection, and communication protocols have a significant impact on underlying communication primitives and, in turn, on parallelized inference serving performance. Recent works both leverage existing networking advancements and investigate more sophisticated strategies for parallelizing LLM deployment, aiming to control the inter-accelerator communication overheads across different components of the system stack~\cite{infiniteHBD, bandPilot}. For example, Ulysses~\cite{DeepSpeed_Ulysses} and RingAttention~\cite{ringAttention} employ sequence parallelism to handle the large KV cache associated with long-context inference by partitioning the input sequence and distributing segments across accelerators. TokenRing~\cite{tokenRing} extends these approaches by effectively overlapping computation and communication, thus overcoming the idleness of compute units and improving their utilization.

Much of the existing literature has focused mainly on optimizing training objectives to alleviate the substantial resource demands associated with {\em training} large-scale models. However, the LLM training procedure involves bidirectional passes, dynamic weight updates, and distinct KV cache requirements, fundamentally differing from {\em inference}. As a result, dedicated efforts are required to design parallelization strategies specifically for LLM inference services. The two studies most relevant to our work are (i) DeepSpeed-Inference~\cite{deepspeedInf}, providing latency and throughput orientation control for the inference services, and (ii) ~\cite{effScalingInfe}, analytically modeling the inference latency and thoroughly analyzing TP deployment on TPU-v4 for long sequences with tight latency constraints. Despite the widespread adoption of LLM-powered applications and various LLM inference services~\cite{deepspeedInf, vLLM, tensorRT_par, sglang}, to our knowledge, the literature has not satisfactorily addressed the limits of latency-flexibility and high-throughput potentials under varying batch sizes, input and model characteristics, and parallelization strategies. Our study aims to fill this gap by extensively investigating the impacts of parallelization preferences, batch sizes, and sequence length variations using representative datasets, as well as diverse model sizes and quantization levels on inference latency and system throughput.

%% file: sections/sim_validation.tex

\section{In-House Simulator and Its Validation} \label{sec:simval} 
In this study, our primary goal is to analyze the latency–flexibility and latency–throughput tradeoffs associated with different batching and parallelization strategies when serving inference for diverse prompts drawn from representative datasets on dense LLMs. Exploring these tradeoffs requires comprehensively evaluating hundreds of combinations of batching policies, input characteristics, model and hardware configurations, and parallelization schemes. To make this exploration feasible, we employ an in-house simulator that estimates key inference latency metrics, e.g., time to first token (TTFT) and time per output token (TPOT), and derives throughput values accordingly. This approach enables systematic, large-scale experimentation within practical time and resource constraints.

\begin{figure*}[htbp!]
    \centering
    \includegraphics[width=0.75\linewidth]{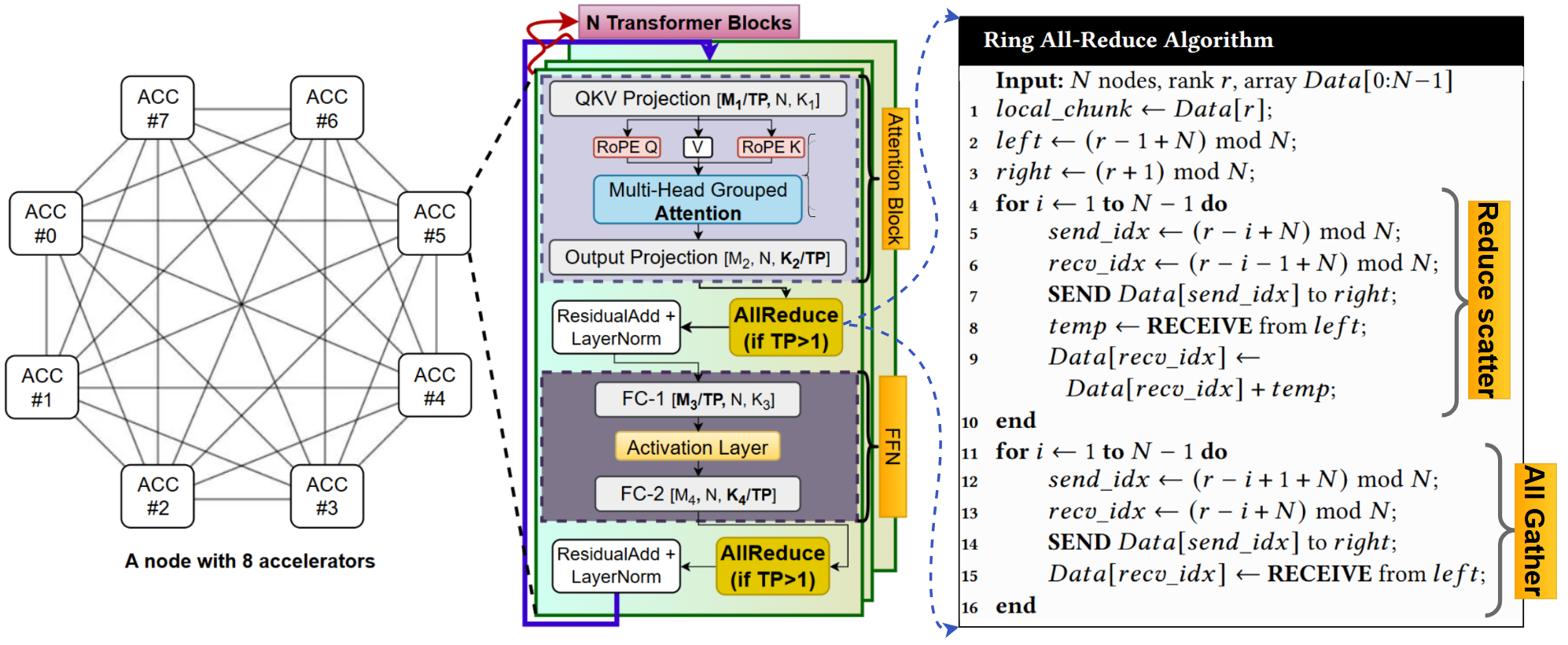}
    \vspace{-2mm}
    \caption{An illustrative overview of the Tensor Parallelism workflow for LLM inference.}
    \label{fig:tp}
    \Description{An illustrative overview of the Tensor Parallelism workflow for LLM inference.}
    \vspace{-2mm}
\end{figure*}

Our simulator analytically models end-to-end inferences of LLM workloads across different system setups, inference models, and input characteristics. More specifically, the simulator models transformer blocks on top of individual representations of kernels, such as Attention and GEMMs, including their state-of-the-art versions (i.e., GQA~\cite{gqa}), with variable hyperparameter dimensions, as in the real model architectures. The simulator achieves a complete representation capability by cascading an equal number of transformer layers to those in the actual LLM. Furthermore, it virtually mimics nodes equipped with multi-GPUs, clusters (i.e., multi-node), the interconnection between both within and across nodes, and the high-level compute and memory hierarchies, encompassing memory bandwidth and read/write latency, to accurately model the target systems under compute- or memory-bound conditions. 

Figure \ref{fig:llama3_70b_main} shows the normalized execution times—relative to MI325x GPU \cite{mi325x}—obtained from our simulator for forward passes through the Llama 3.1–70B FP8 transformer. The results span a range of input sequence lengths (ISL), output sequence lengths (OSL), and batch sizes (BS), as indicated on the x-axis. As shown, the simulator accurately captures the transformer execution on MI325x GPUs for the prefill and decode phases, illustrated by the orange and green bars, with accuracies greater than 90\% and 86\%, respectively. Furthermore, Figure~\ref{fig:llama3_70b_break} reveals the contribution of individual kernels to transformer execution, along with their errors relative to the total transformer pass time between the simulator and actual hardware. Although the simulator may exhibit minor inaccuracies in predicting individual kernel runtimes, these deviations are not critical for our purposes. What matters is that the simulator reliably captures the overall latency and throughput {\em trends} across different batching strategies, input characteristics, and system configurations, which is the primary focus and requirement of our study.

To extend the validation of our simulator for a larger-scale model running on multiple GPUs, we conduct a similar correlation analysis for the Llama 3.1-405B FP8 model, configured with \textit{TP8} parallelization (i.e., tensor parallelism with a degree of 8). Figures~\ref{fig:llama3_405b_main} and~\ref{fig:llama3_405b_break} present the normalized transformer and kernel-wide execution comparisons between the simulator and a node setup equipped with eight MI325x GPUs, for both the prefill and decode phases under various ISL, OSL, and batching configurations. The simulator reproduces the prefill execution behavior on hardware for four different prefill lengths with an average error of 17.25\%, which is slightly higher than the error observed in single-GPU validation. Moreover, the simulation accuracy for the decode phase is comparatively lower than that of prefills, reflecting the increased complexity introduced by deep parallelization. The slight accuracy drop introduced with a deep parallelization scheme can be attributed to dynamic voltage-frequency variations among GPUs, causing diverse kernel executions that delay all-reduce synchronization, as well as some suboptimal kernel performances on the actual hardware. Furthermore, FC-1, FC-2, and all-reduce kernels dominate transformer execution in both inference phases, and the combined accuracy deviations of these dominant kernels negatively contribute to the correlation for end-to-end executions on the simulator and silicon. For instance, the increasing error margin while predicting the decode inference passes through the Llama 3.1-405B TP8 setup is mainly caused by the accumulated error influence of FC-1 and two all-reduce kernels. Still, as mentioned above, our simulator is very successful in identifying the throughput and latency trends for various parallelism strategies, batching configurations, diverse input conditions, and distinct model features.


%% file: sections/methods.tex

\section{Parallelization Strategies} \label{sec:parStrategies}
This section provides an in-depth analysis of the parallelization strategies investigated in our work and elucidates the structural mapping of transformer layers, along with accelerators, within a node environment when LLMs are deployed with these strategies.

\begin{figure*}[t]
    \centering
    \includegraphics[width=0.75\linewidth]{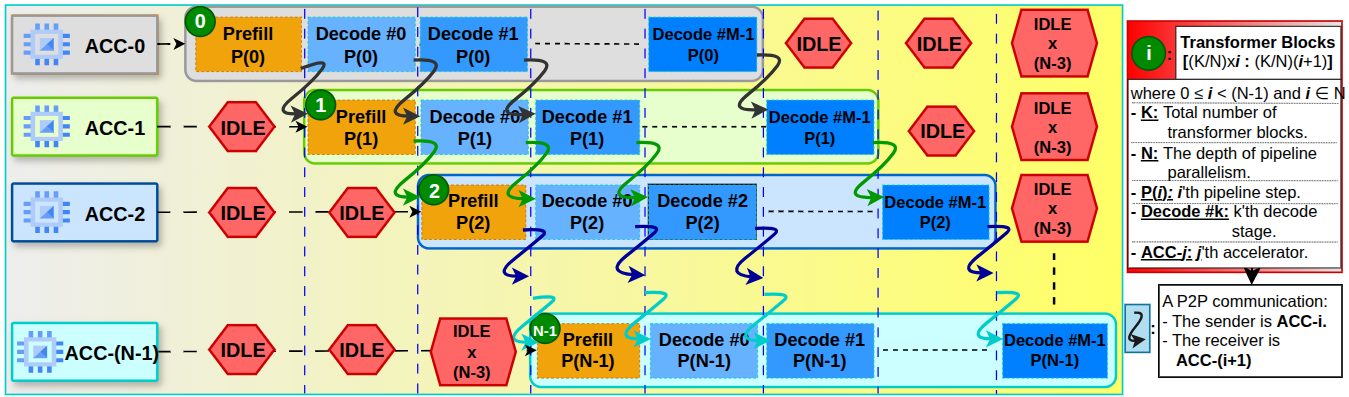}
    \caption{A schematic overview of the Pipeline Parallelism and its workflow distribution for LLM Inference.}
    \vspace{-2mm}
    \Description{A schematic overview of the Pipeline Parallelism and its workflow distribution for LLM Inference.}
    \label{fig:pp}
    \vspace{-2mm}
\end{figure*}

\subsection{Tensor Parallelism} 
Figure~\ref{fig:tp} demonstrates an overview of the Tensor Parallelism (TP) workflow in a node equipped with 8 accelerators and a mesh interconnect, providing peer-to-peer (P2P) connections. TP splits the model weights and attention heads uniformly, distributing the KV cache entries across accelerators, and thereby allows accommodating larger batches and inference through models that do not fit into a single accelerator memory. In transformer blocks, GEMM operations, such as FC-1 and FC-2, with dimensions $[M, N, K]$ compute $C = \alpha AB + \beta C$, where $\alpha \text{ and } \beta$ are scalar coefficients, $A \in R^{M\times K} \text{ and } B \in R^{K\times N}$ are input matrices, and the output matrix is $C \in R^{M\times N}$. To demonstrate a typical tensor sharding in TP, we divide QKV Projection and FC-1 GEMMs, each with unique dimensions $M, K$, \textit{row-wise}, where $N$ corresponds to the prefill length or batch size in the prefill and decode phases, respectively. In this scheme, each accelerator executes QKV Projection and FC-1 layers with dimensions $[M_1/TP, N, K_1]$ and $[M_3/TP, N, K_3]$, respectively, where $TP$ represents the depth of TP. Similarly, we shard Output Projection and FC-2 GEMMs \textit{column-wise}, resulting in dimensions $[M_2, N, K_2/TP]$ and $[M_4, N, K_4/TP]$, respectively, as illustrated in the transformer blocks of Figure~\ref{fig:tp}.

As previously stated, the system evenly splits attention heads along with participant accelerators. For example, $TP=4$ divides the 64 attention heads of the Llama 3.1-70B model into four groups, and the infrastructure assigns the first group, comprising 16 heads, to ACC-0 and the remaining 48 heads to ACC-1, ACC-2, and ACC-3, respectively. TP requires aggregating partial outputs of attention and FFN blocks on each accelerator via {\it all-reduce} operations before the residual addition layers. \textit{Ring algorithm}, commonly preferred method performing the all-reduce operation for TP as illustrated in Figure~\ref{fig:tp}, consists of two key phases--\textit{reduce-scatter} and \textit{all-gather}~\cite{allGather}--which together handle the inter-accelerator communication through \textit{send} \& \textit{receive} operations and perform the necessary summation (line 9). Accordingly, the accelerator group participating in TP executes the forward pass on a transformer block synchronously to generate output tokens for a batch of requests, as depicted in the middle subfigure.

For a number of batched requests, even though TP incurs the communication overhead of all-reduce layers, splitting attention and FFN blocks across accelerators provides a faster forward pass through a transformer block. For the prefill phase, which is a compute-bound workload~\cite{distServe}, increasing the TP size shortens the execution of core transformer layers, as larger TP schemes engage more accelerators and consequently dedicate more compute resources to execute an equivalent amount of workload. Additionally, the memory space dedicated to the model parameters is reduced as a result of sharding, and the saved memory can be utilized for larger KV caches to form larger batches during inference. TP thus addresses the memory-bound nature of the decode phase, which can potentially help to benefit more from the compute units. However, the more accelerators participating in inter-device communication in wider TP schemes will result in more units to be included in all-reduce operations, which can potentially limit the latency benefit of widening the TP. Even though a larger-scale TP deployment for a given batch size results in fewer data to be transmitted with each send-and-receive step, due to a smaller partitioning of GEMM and attention workloads, it leads to more loop iterations (line 4 and line 11) for both reduce-scatter and all-gather steps, resulting in more communication steps. Therefore, the depth of parallelization, the underlying all-reduce algorithm, batching, interconnect topology, and link bandwidth are key factors in determining the limits of achievable latency while serving inferences for diverse models and input prompts. 

Despite the fact that TP offers latency flexibility for inference-serving systems, its throughput (output tokens per second) impact can vary depending on the model size and the depth of parallelization. To illustrate, a system deploying the Llama 3.1-405B FP8 model with a parameter size of 405GB has a KV cache of $2\text{ x }(2\text{ x }256\text{GB} - 405\text{GB}) \approx 214\text{GB}$ for two data-parallel $TP=2$ systems and $4\text{ x } 256\text{GB} - 405\text{GB} \approx 619\text{GB}$ for the $TP=4$ system, where systems consist mainly of four MI325x GPUs, each with 256GB of device memory. Under the stated conditions, the $TP=4$ setup provides a KV cache capacity 2.89 times that of two-fold data-parallel $TP=2$ systems, while experiencing six send-and-receive steps (i.e., three for each reduce-scatter and all-gather), four more compared to $TP=2$. Furthermore, the $TP=4$ system, with a much larger batch size, migrates a larger amount of data at each communication step than the $TP=2$ system when targeting high throughput, resulting in a significant all-reduce latency. Therefore, the size of the model, the degree of parallelism, as well as the input characteristics determine the KV cache behavior, and they are key to achieving a system designed to satisfy latency and throughput demands. In our evaluations, we empirically analyze the latency flexibility and the latency-throughput interplay for various sizes of $TP$ within a node environment. 

\subsection{Pipeline Parallelism} 
Unlike TP, which performs intra-layer sharding alongside transformer layers, such as GEMM and attention kernels, Pipeline Parallelism (PP) distributes transformer blocks among accelerators (i.e., inter-layer split) based on the depth of parallelization. Figure~\ref{fig:pp} demonstrates a high-level overview of PP deployment for LLM inference. Suppose that a system with the $PP=N$ configuration serves inferences through $K$ transformers for a dense model, where the transformer blocks are divided into sets of $K/N$, and each is mapped to an accelerator. Across all $N$ pipeline stages, each pipeline stage $P(i)$--running on ACC-$i$ for $i \in [0, N-1]$ as represented in Figure~\ref{fig:pp}--is responsible for processing requests across transformer layers in the range $[(K/N)\text{ x }i:(K/N)\text{ x }(i+1)]$, where the annotations describing the range of transformer layers assigned to pipeline stages are on the right side of the illustration. Both prefill and decode begin to perform inference on stage $P(0)$, running on ACC-$0$, by processing the inference workload through transformer blocks in the range $[0:K/N]$. Following completion of stage $P(0)$, the serving framework transmits activations of the $K/N$'th block from ACC-$0$ to ACC-$1$, as illustrated by the leftmost black arrow directed from $P(0)$ to $P(1)$ in Figure~\ref{fig:pp}, and inference continues on the pipeline stage $P(1)$ by passing through the transformer layers, spanning $[K/N:(K/N)\text{ x }2]$ interval. While $P(1)$ performs the operation on the batch completed in $P(0)$, $P(0)$ starts either processing a new batch of requests or generating the next token for the batch that has already passed through the $P(N-1)$ step. Following this procedure, PP enables the inference system to execute a parallel pipelined execution of multiple batches simultaneously.

\begin{figure*}[t]
    \begin{subfigure}{0.35\textwidth}
        \centering
        \includegraphics[width=\linewidth]{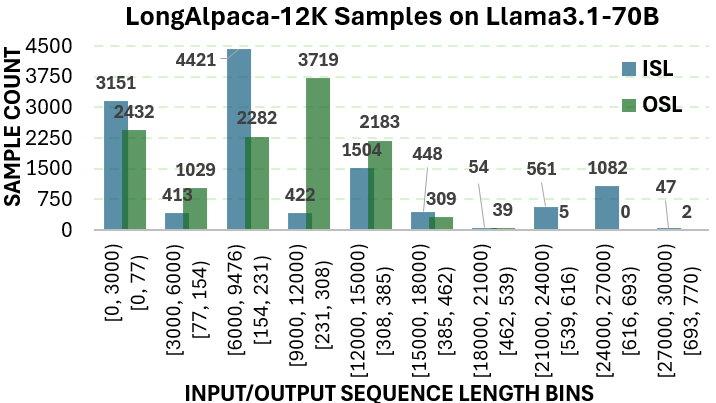}
        \caption{LongAlpaca dataset with 12K samples on Llama 3.1-70B model.}
    \end{subfigure}
    \hfill 
    \begin{subfigure}{0.33\textwidth}
        \centering
        \includegraphics[width=\linewidth]{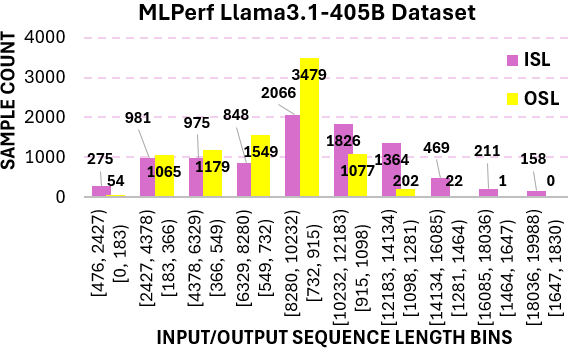}
        \caption{MLPerf dataset with 8313 samples on Llama 3.1-405B model.}
    \end{subfigure}
    \hfill 
    \begin{subfigure}{0.3\textwidth}
        \centering
        \includegraphics[width=\linewidth]{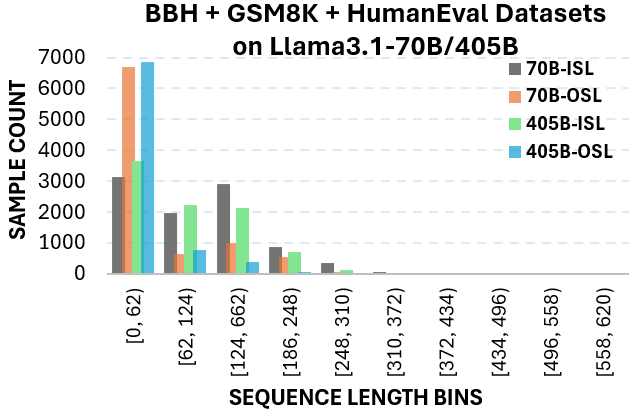}
        \caption{Combined BBH, GSM8K, and HumanEval datasets on both models.}
    \end{subfigure}
    \caption{Input and output sequence length distribution for LongAlpaca, MLPerf, and the combination of BSH, GSM8K, and HumanEval datasets on Llama 3.1-70B and -405B models.}
    \label{fig:datasets}
    \Description{Input and output sequence length distribution for LongAlpaca, MLPerf, and a combination of BSH, GSM8K, and HumanEval datasets on Llama 3.1-70B and -405B models.}
    \vspace{-3mm}
\end{figure*}

Even though the PP scheme decreases the number of transformer blocks assigned to accelerators participating in parallelism, it has no effect on the execution time per transformer pass, in contrast to TP, as the compute resources allocated to execute one transformer block remain unchanged. Instead, it alleviates memory pressure per accelerator due to the distributed structure of transformer layers, allowing larger KV caches and forming larger batches per accelerator, and further enables the simultaneous processing of multiple batches at inference time. Whereas deeper PP setups experience more P2P communication between pipeline stages, the activation data transmitted across peers is the same when the nano-batches (i.e., a batch of requests per accelerator), despite containing distinct requests, are equal in size, resulting in identical send-and-receive overhead across different P2P data transactions. Consequently, the mitigated memory pressure via PP, with distributing transformer layers along the pipeline stages, fails to yield any latency benefit. Additionally, deeper PP schemes can negatively contribute to the latency objective of inference serving, as they expose increased P2P communication across a wider range of pipeline stages. 

Still, distributing the transformer blocks and their set of parameters across accelerators via PP potentially leads to throughput improvement by allowing larger nano-batches that concurrently run on distinct pipeline stages and result in a larger global batch (i.e., the total batches the system actively processes). To illustrate, each accelerator manages \textasciitilde202.5GB and \textasciitilde101.25GB of the model parameters when serving the Llama 3.1-405B FP8 model under $PP=2$ and $PP=4$ settings, respectively. Hence, each accelerator has $256\text{GB} - 202.5\text{GB} \approx 53.5\text{GB}$ and $256\text{GB} - 101.25\text{GB} \approx 154.75\text{GB}$ available space for the KV cache when configured with the $PP=2$ and $PP=4$  settings, respectively, where both systems run on four MI325x GPUs. That is, the $PP=4$ scheme offers to store a 2.89 times larger KV cache, which leads to significantly larger nano-batches being operated with compared to the $PP=2$ scheme when deployed through four MI325x GPUs. Consequently, deeper PP schemes enable systems to operate with larger batches, which can positively contribute to overall system throughput by eliminating the limited compute utilization throughout memory-bounded decode phases. Moreover, the P2P communication overhead introduced by PP is typically much smaller than the impact of all-reduce kernels introduced by TP, since transfers between pipeline stages--each less costly than a single all-reduce step--occur $PP_{depth} - 1$ times between transformers while all-reduce kernels occur twice the total count of transformer blocks. As a result, employing PP for serving dense LLMs potentially increases the overall latency due to substantially larger global batch sizes and overhead of P2P communications, but improves system throughput. In what follows, we demystify the latency–throughput trade-off under varying depths of PP within a node and provide a comparative analysis of PP and TP.

\subsection{Hybrid Tensor–Pipeline Parallelism}
Hybrid PP and TP deployments distribute the transformer blocks across sets of accelerators, where each set is responsible for one pipeline stage, and the depth of PP determines the total number of sets. Within each pipeline stage, a segment of the transformer layers, whose size is equivalent to the total transformer layers divided by the depth of PP, is split among the accelerators involved in TP. For example, when the Llama 3.1-405B model is configured using the $TP=2, PP=2$ parallelization setting, PP stages--$P(0) \text{ and } P(1)$, each running on two accelerators--compute the inference through 63 transformer blocks where the model consists of 126 transformer blocks, as mentioned in Table~\ref{tab:model_specs}. Furthermore, the $TP=2$ setup within each $P(i)$ with $i\in\{0,1\}$ splits the internal transformer kernels and their respective tensors into two portions such that an accelerator within each $P(i)$ stage computes half of the internal transformer kernels, such as calculating the QKV Projection GEMM with dimensions of $[M_1/2, N, K_1]$, but for only 63 transformer blocks. Accordingly, the hybrid usage of PP and TP is capable of potentially achieving high throughput for increasing the depth of PP while maintaining low inference latency along with the size of TP. We analyze the latency–throughput tradeoff under varying hybrid TP and PP configurations in Section~\ref{sec:evaluations}.   


%% file: sections/evaluations.tex
\section{Experimental Evaluations} \label{sec:evaluations}
\subsection{Experiment Setup}
{\bf Models:} Llama 3.1~\cite{llama3} models with 8B, 70B, and 405B versions and context length support of up to 128k tokens are claimed to be competitive in general knowledge and reasoning tasks, requiring long contextual memory, such as summarization and coding~\cite{longContSummary}. As explained in Table~\ref{tab:model_specs}, the 70B and 405B variants have 64 and 128 query heads and both employ grouped 8 KV heads (i.e., GQA~\cite{gqa}), where the head dimension is 128, resulting in hidden dimension sizes of 8192 and 16384, respectively. The transformer networks, whose high-level structure is shown in Figure~\ref{fig:tp}, consist of 80 and 126 blocks for the 70B and 405B versions, respectively. 

\noindent {\bf Datasets:} Considering that the Llama 3.1-70B and -405B models typically serve inferences for a wide range of context lengths, we perform inference simulations with datasets representing both long and short input behaviors. The combination of \textit{BIG-Bench Hard} (BBH)~\cite{bbhData}, \textit{Human Evaluation}~\cite{humanEvalData}, and \textit{GSM8K}~\cite{gsm8kData} datasets involves comparably shorter sequences when interacting with both models. The BBH dataset presents samples to analyze advanced reasoning, comprehension, and multi-step logic capabilities, and GSM8K is used to measure the multi-step arithmetic and reasoning success of LLMs through high-quality grade-school math problems. Human Evaluation presents prompts to assess the functional correctness of programming tasks. 

On the other hand, \textit{LongAlpaca}~\cite{longAlpaca} and \textit{MLPerf}~\cite{mlperf} datasets bring longer sequences during inferences through the Llama 3.1-70B and -405B models, respectively. LongAlpaca is utilized to evaluate model capabilities on long-context reasoning and instruction-following tasks, or to fine-tune the models. MLPerf dataset, with carefully selected 8313 prompts from the Longbench~\cite{longbench}, GovReport~\cite{govReport}, and Ruler~\cite{ruler} datasets, serves as a cornerstone benchmark for inference evaluations. LongAlpaca and MLPerf inference datasets with longer prompts are created to test the context scaling capabilities of models, which present larger context windows, for real-world tasks such as summarization and code understanding. Figure~\ref{fig:datasets} illustrates the distribution of the sequence length characteristics of each dataset, where ISL and OSL stand for the input and output sequence lengths, when their inferences are served through the target models.

\begin{figure*}[!htb]
    \centering
    \begin{subfigure}{0.485\textwidth}
        \centering
        \includegraphics[width=\textwidth]{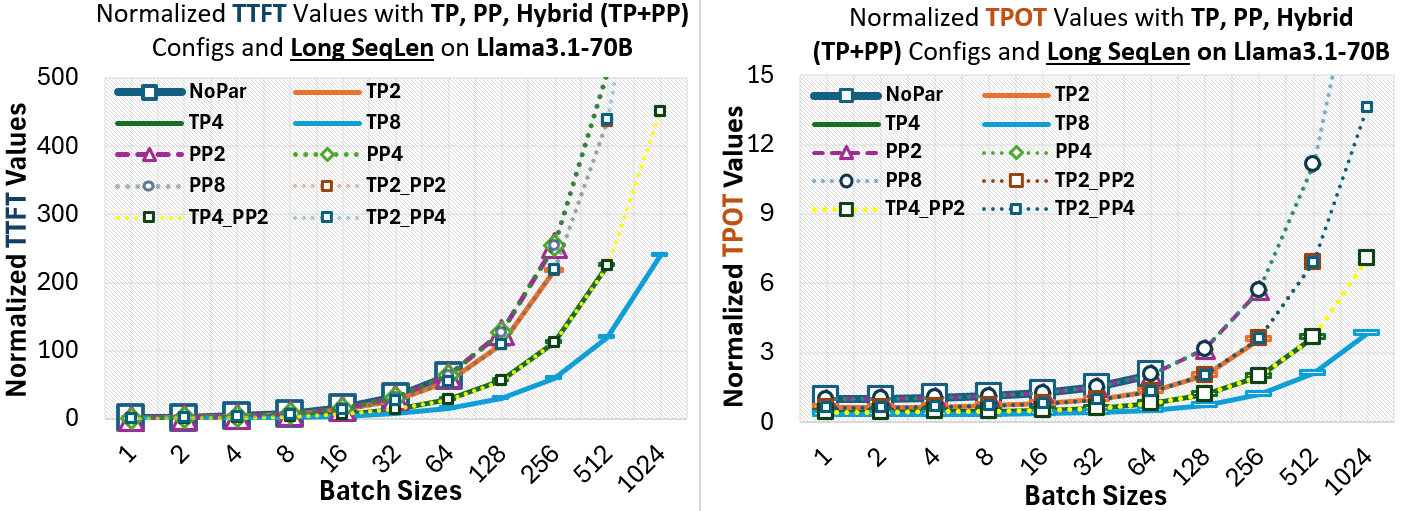}
        \caption{For the LongAlpaca dataset.}
        \label{fig:llama3_70b_lsl}
    \end{subfigure}
    \hfill
    \begin{subfigure}{0.49\textwidth}
        \centering
        \includegraphics[width=\textwidth]{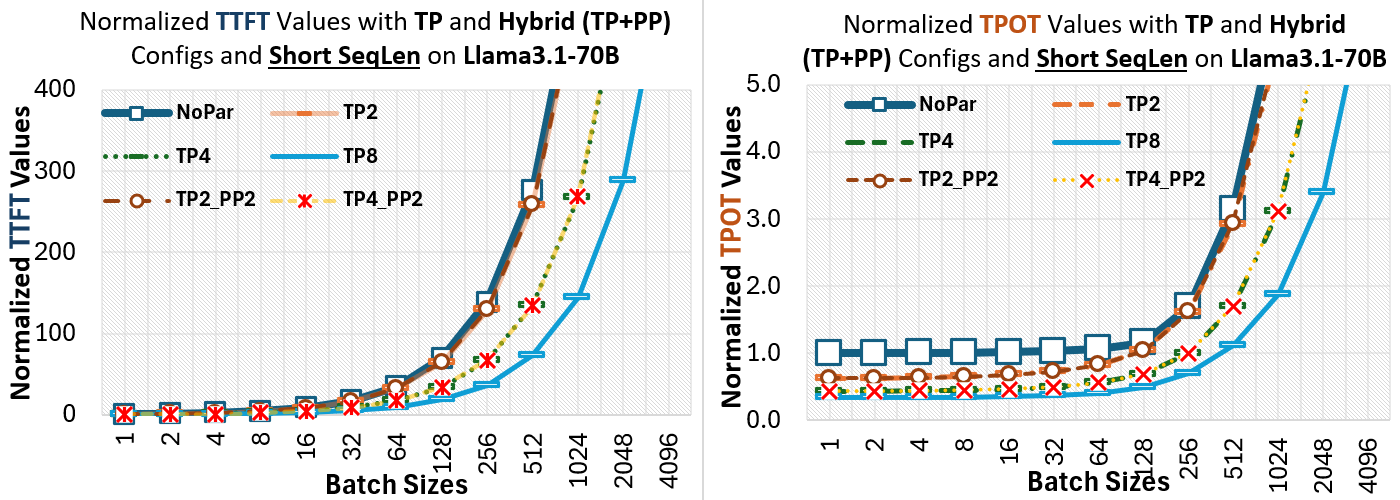}
        \caption{For the BBH, GSM8K, and Human Evaluation datasets.}
        \label{fig:llama3_70b_ssl}
    \end{subfigure}
    \vspace{-2mm}
    \label{fig:llama370b_lat}
    \caption{Normalized TTFT and TPOT values when serving inferences for datasets in (a) and (b) through Llama 3.1-70B.}
    \Description{Normalized TTFT and TPOT values when serving inferences through Llama 3.1-70B.}
    \vspace{-2mm}
\end{figure*}

\begin{table}[h]
  \centering
  \caption{Average input and output sequence lengths of the evaluation datasets when their inferences are served through target LLMs.\label{tab:datasets}}
  \vspace{-2mm}
  \begin{tabular}{p{0.08cm} p{5cm} | p{0.45cm} | p{0.45cm} | p{0.75cm}}
  \toprule
  \multicolumn{2}{c}{\textbf{Datasets}} & ISL & OSL & TOTAL\\
  \midrule
  \multirow{2}{*}{\rotatebox{90}{\textbf{70B}}} & \small LongAlpaca~\cite{longAlpaca} & 9092 & 208 & \textbf{9350} \\
  & \small BBH, GSM8K, Human Evaluation~\cite{bbhData, gsm8kData, humanEvalData} & 106 & 26 & 132 \\
  \midrule
  \multirow{2}{*}{\rotatebox{90}{\textbf{405B}}} & \small MLPerf~\cite{mlperf} & 9428 & 684 & \textbf{10112} \\
  & \small BBH, GSM8K, Human Evaluation & 89 & 20 & 110 \\
  \bottomrule
  \end{tabular}
  \vspace{-2mm}
\end{table}

We experiment with LongAlpaca and the combination of BBH, Human Evaluation, and GSM8K datasets to analyze parallelization performance while serving inference through the Llama 3.1-70B model under both long- and short-sequence input characteristics. For a similar analysis but with a larger model, we use MLPerf and the same combined datasets to determine latency and throughput behaviors under various Llama 3.1-405B inference setups. In our experiments, we plug in average ISL and OSL numbers for each individual dataset, as outlined in Table~\ref{tab:datasets}, to introduce representative input characteristics for our latency and throughput analysis.  

\begin{table}[!htbp]
  \centering
  \caption{GPU specifications.\label{tab:gpu_specs}}
  \vspace{-2mm}
  \begin{tabular}{p{1.5cm} | p{0.75cm} | p{1.2cm} | p{1.5cm} | p{1.5cm}}
  \toprule
  \textbf{GPU} & HBM Mem. & Arch. & Precision Support & Inter Connect BW \\
  \midrule
  \textbf{MI325x} ~\cite{mi325x} & 256GB & CDNA-3 & \small 64, 32, 16, and 8-bit & \small 128GB/s bidirect \\ 
  \textbf{MI355x} ~\cite{mi350x} & 288GB & CDNA-4 & \small 64, 32, 16, 8 and 4-bit & \small 153.6GB/s bidirect \\
  \bottomrule
  \end{tabular}
  \vspace{-1mm}
\end{table}

\noindent {\bf Inference Setup:} Our simulations mimic AMD's Instinct GPU-node platforms equipped with eight MI325x~\cite{mi325x} or MI355x~\cite{mi350x} GPUs and their all-to-all interconnect between GPU pairs. Table~\ref{tab:gpu_specs} summarizes the key specifications of our target platforms, particularly the GPU HBM capacities and precision supports. Additional details on architectural topology and compute units are available in the corresponding reference datasheet. 

\noindent {\bf Evaluation Metrics:} The broad spectrum of LLM-powered applications entails diverse performance objectives, ranging from stringent latency requirements to balanced latency-throughput tradeoffs and pure throughput optimization, depending on whether the workload involves real-time, interactive multi-user tasks or offline batch processing~\cite{mlcommons}. Accordingly, the time-to-first-token (TTFT) and time-per-output-token (TPOT) metrics are used to evaluate inference latency, while the total output tokens per second (TPS) metric is preferred to evaluate throughput. In this regard, our experiments quantify the latency and throughput for various parallelization and batching configurations while simulating inferences for 8-bit Llama 3.1-70B and 4-bit Llama 3.1-405B quantized models on nodes equipped with MI325x and MI355x GPUs, respectively, through the datasets listed in Table~\ref{tab:datasets}.  

\subsection{Experiment Results} 
\subsubsection{Latency Flexibility Analysis} \hspace{0pt}\\[2pt]
{\bf Llama 3.1-70B.} Figures~\ref{fig:llama3_70b_lsl} and~\ref{fig:llama3_70b_ssl} plot the TTFT and TPOT values when serving inferences through distinct Llama 3.1-70B deployment setups, such as parallelization schemes within a node equipped with MI325x GPUs, for two datasets representing long and short sequences. As noted in Section~\ref{sec:simval}, since the exact numbers may slightly vary depending on the inference serving software, system settings, and kernel implementations, we normalize the TTFT and TPOT values with respect to the non-parallel case, labeled as NoPar, with a batch size of 1, to analyze the trends for latency objectives. As Figure~\ref{fig:llama3_70b_lsl} shows, TP8 (that is, $TP=8$) outperforms other possible parallelization settings on both TTFT and TPOT latency metrics for all batch sizes. Similarly, TP4 and $TP4\_PP2$ (that is, $TP=4$ and $PP=2$), each shallower parallelization than TP8, achieve smaller TTFT and TPOT values relative to all only-PP, hybrid TP \& PP with lower degrees of TP, and without parallel setups. Therefore, increasing the degree of TP, such as from TP4 to TP8, enhances the latency flexibility. For example, when the batch size is set to 256 while serving for the LongAlpaca dataset, TP8 achieves (1.87×, 1.67×) and (3.61×, 3.01×) faster (prefill, decode) executions than TP4 and TP2, respectively, where setups using a parallelization depth below 2 cannot form a batch of 256 prompts as the KV cache exceeds the GPU memory. Sharding Attention and FFN blocks within transformers across TP-participating GPUs yields dedicating more compute units to each transformer pass, leading to faster inferences, especially for the prefill phase, as it is mainly compute-bounded. However, PP, which distributes the sequentially executed transformer blocks across GPUs in an inter-layer fashion, enables multiple batches to be processed concurrently rather than devoting more compute resources to an individual transformer pass; thus, it fails to reduce TTFT and TPOT latencies.

The batching preference and prompt sequence lengths affect the TTFT and TPOT latencies as well as the parallelization. The prefill and decode workload sizes—determining their execution durations—depend on the prefill length, batch size, and model itself, where the prefill length corresponds to the batch size $\times$ ISL, and the decode workload size is mainly determined by the batch size. The dimension $N$ in transformer GEMMs, as illustrated in Figure~\ref{fig:tp}, corresponds to the prefill length and batch size for the prefill and decode phases, respectively. For example, small batch sizes directly lead to a reduced workload for both prefill and decode transformer passes, and result in shorter TTFT and TPOT accordingly, as both latencies decrease to the left on the x-axis of Figures~\ref{fig:llama3_70b_lsl} and~\ref{fig:llama3_70b_ssl}. Moreover, short input prompts corresponding to small ISL values decrease the prefill workload and reduce TTFT latency. For instance, serving inference for the combined BBH, Human Evaluation, and GSM8K datasets within a TP8 setup and a batch size of 256 achieves the first token (i.e., prefill) 1.63 times faster than serving for prompts from the LongAlpaca dataset. However, neither ISL nor OSL directly influences the TPOT values, since the decode phase generates a number of tokens equal to the batch size at each step. The indirect way that prompt sequence lengths affect TPOT is through memory access delays: longer prompts require larger KV caches to be loaded onto the cores, even when operating with the same batch size, but their impact on TPOT would still be relatively small. 

\begin{figure}[h]
    \centering
    \includegraphics[width=1\linewidth]{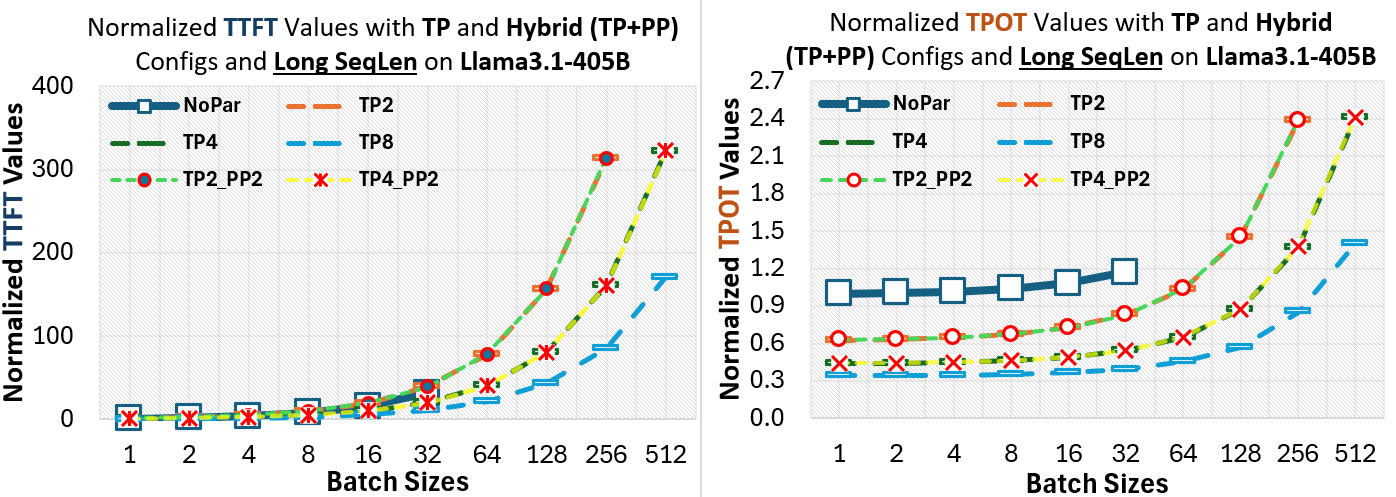}
    \vspace{-1mm}
    \caption{Normalized TTFT and TPOT values for Llama 3.1-405B inference serving with varying parallelization techniques \& depths, and batch sizes.}
    \Description{Normalized TTFT and TPOT values for Llama 3.1-405B inference serving with varying parallelization techniques \& depths, and batch sizes.}
    \vspace{-2mm}
    \label{fig:llama3_405B_lat}
\end{figure}

\noindent {\bf Llama 3.1-405B.} To investigate the latency flexibility for a larger model, where significantly less memory is available for the KV cache compared to smaller models, we extend our exploration to serving inferences through the Llama 3.1-405B model with prompts from the MLPerf inference dataset. The size of 405B parameters at 4-bit quant, \textasciitilde202.5GB, can fit into one MI355x GPU memory, and \textasciitilde85GB space remains available for the KV cache. Since only-PP schemes fail to reduce TTFT and TPOT latencies, as explored through Llama 3.1-70B analysis, we share the latency metrics for TP, hybrid TP \& PP, and without parallelization configurations, as shown in Figure~\ref{fig:llama3_405B_lat}. Consistent with the takeaways from the latency analysis of Llama3.1-70B inference, deeper TP schemes, such as TP4 or TP8, drastically reduce the TTFT and TPOT values for inferences through the 405B model. For a batch size of 256, requiring the overall parallelization depth to be at least 2 to accommodate the corresponding KV caches, TP8 completes (prefill, decode) phases (1.89×, 1.61×), (1.90×, 1.62×), and (3.67×, 2.81×) faster than TP4, TP4\_PP2, and TP2, respectively. The underlying reason that the TP4 setting provides slightly better latency than the hybrid TP4\_PP2 configuration is due to the delay incurred by P2P communications in PP. Moreover, the larger model weights leave limited space for KV caches, thus restricting the maximum allowable batch size and ultimately reducing compute utilization. Hence, the 405B model, with more transformer blocks and larger tensor kernels than the 70B version, exhibits longer TTFT and TPOT for the same batch size and similar sequence lengths.

\begin{figure}[!htbp]
    \centering
    \begin{subfigure}{0.235\textwidth}
        \centering
        \includegraphics[width=\textwidth]{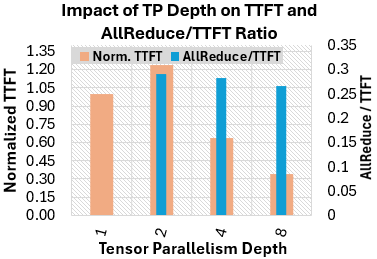}
        \caption{The tradeoff between TP size and TTFT \& all-reduce latencies.}
        \label{fig:tp_depth}
    \end{subfigure}
    \hfill
    \begin{subfigure}{0.225\textwidth}
        \centering
        \includegraphics[width=\textwidth]{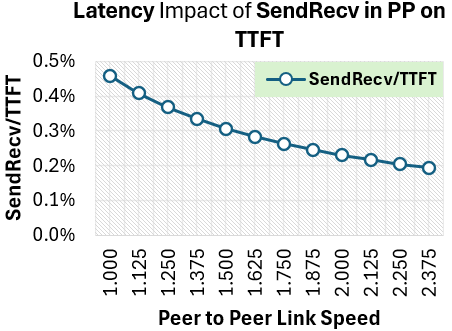}
        \caption{The tradeoff between the link speed and TTFT latency in PP.}
        \label{fig:pp_sendRecv}
    \end{subfigure}
    \hfill
    \begin{subfigure}{0.3\textwidth}
        \vspace{1.5mm}
        \includegraphics[width=\linewidth]{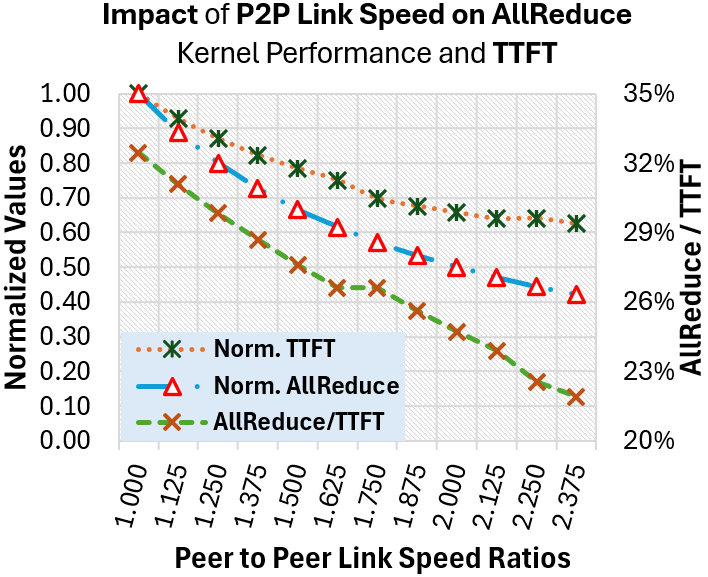}
        \caption{The relationship between the link speed and TTFT \& all-reduce latencies.}
        \label{fig:ttftAllreduce}
    \end{subfigure}
    \label{fig:tpDepthAllReduce}
    \caption{Impact of TP size and interconnection link speed on TTFT, all-reduce and P2P communication durations in TP and PP.}
    \Description{Impact of interconnection link speed on the TTFT latency in PP and TP.}
    \vspace{-3mm}
\end{figure}

\noindent {\bf Latency Overheads of Intra-Node Communications.} 
To study the latency overhead of intra-node communication for varying TP sizes, we measure TTFT and all-reduce durations for possible TP settings in a node when serving inferences for the MLPerf dataset through the Llama 3.1-405B model with a batch size of 32. Figure~\ref{fig:tp_depth} demonstrates the normalized TTFT and all-reduce durations relative to the TP1 (non-parallelized) case and the TTFT latency of the corresponding TP scheme, respectively. The TP8 and TP4 schemes reduce the TTFT durations by 68\% and 38\%, respectively, but the overhead of all-reduce operations causes TP2 to process prefill slower than TP1. Although deeper TP schemes accelerate all-reduce kernels by improving the aggregate active bandwidth capacity~\cite{aggregateLink}, all-reduce-to-TTFT ratios remain almost unchanged, indicating that they introduce an inevitable latency overhead regardless of the TP size. Additionally, to analyze the impact of link speed on all-reduce and TTFT, we sweep the aggregate link speed from 256 GB/s to 608 GB/s for serving inferences with the same model, dataset, and batch size, but with the TP8 setup. Figure~\ref{fig:ttftAllreduce} presents the normalized TTFT and all-reduce times (left Y-axis) and the all-reduce-to-TTFT ratio (right Y-axis) under varying aggregate link speeds. We observe that doubling the link speed reduces the TTFT by 34\% and the all-reduce-to-TTFT ratio by 7.7\%. Therefore, improving TTFT and TPOT latencies in deeper TP configurations depends not only on faster interconnects but also on the availability of sufficient compute resources.

Lastly, we investigate the latency impact of communication across inter-pipeline stages, occurring through send-and-receive operations. Figure~\ref{fig:pp_sendRecv} shows the ratios of P2P communications to TTFT while serving inferences for the MLPerf dataset using the PP8-configured 405B model with a batch size of 512. We observe that even for a P2P link speed of 32 GB/s (i.e., one-fifth of the datasheet value), the latency overhead of communications on TTFT is less than 0.5\%. As the total number of inter-pipeline transmissions in PP is equal to $PP_{depth}$ - 1 times, such as 7 times for PP8, and thus significantly fewer than the total all-reduce operations, their latency impact on both TTFT and TPOT is quite small. This also explains the quite similar trends in TTFT and TPOT for PP2, PP4, PP8, and without parallelization schemes in our Llama 3.1-70B and -405B latency analysis. Considering all target physical node systems currently employ all-to-all interconnects for GPU communication, our investigation of interconnect effects on latency focuses solely on link bandwidth and parallelization depth. We therefore do not extend our analysis to how different network topologies might influence parallelization performance. 

\subsubsection{Latency-Throughput Interplay and Throughput Trends} \hspace{0pt}\\[2pt]

\begin{figure}[h]
    \centering
    \begin{subfigure}{0.48\textwidth}
        \centering
        \includegraphics[width=\textwidth]{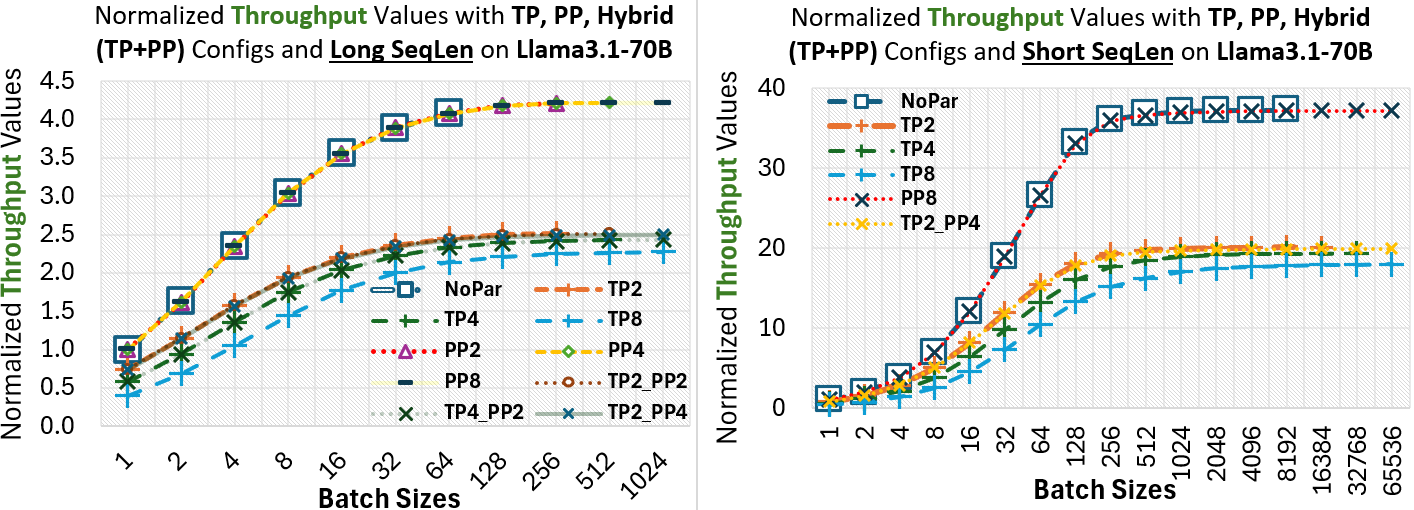}
        \caption{Output TPS serving inferences for LongAlpaca and the combination of the BBH, GSM8K, Human Evaluation datasets, on the left and right, respectively.}
        \label{fig:llama3_70b_tput}
    \end{subfigure}
    \hfill
    \begin{subfigure}{0.48\textwidth}
        \centering
        \includegraphics[width=\textwidth]{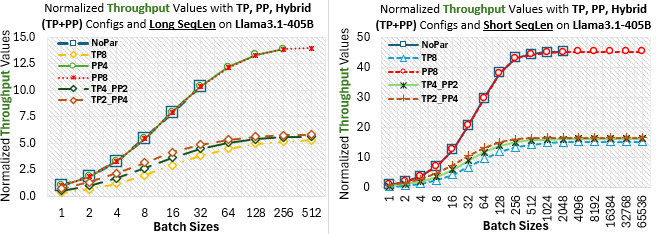}
        \caption{Output TPS while serving inferences for MLPerf and the combination of the BBH, GSM8K, Human Evaluation datasets, on the left and right, respectively.}
        \label{fig:llama3_405b_tput}
    \end{subfigure}
    \label{fig:tput}
    \vspace{-6mm}
    \caption{Normalized output TPS values for inference serving through Llama 3.1-70B (a) and -405B (b) models.}
    \Description{Normalized throughput (output TPS) values for inference serving through Llama 3.1-70B (a) and 405B (b) models.}
    \vspace{-8mm}
\end{figure}

Figures~\ref{fig:llama3_70b_tput} and~\ref{fig:llama3_405b_tput} plot the TPS values achieved with inference systems running the Llama3.1-70B/-405B models within nodes equipped with MI325x/MI355x GPUs, respectively, when serving for the LongAlpaca and MLPerf, and the combined BBH, GSM8K, and Human Evaluation datasets. We apply the formula below to derive the TPS values based on the latency measurements analyzed in the previous section.

{
\setlength{\abovedisplayskip}{-2pt}
    \begin{align*}
    \text{TPS} =
        \frac{\text{Total Output Tokens}}{\text{Total Time (s)}} =
        \frac{G_{\!BS} \times OSL \times N_{\!DP-System}} {Lat_{\!Pref} + OSL \times Lat_{\!Dec}}
    \label{eq:tps} \\
    \text{where} \quad
    \begin{aligned}[t]
        G_{\!BS}   &:\; \text{Global batch size} \\[-2pt]
        OSL &:\; \text{Output sequence length} \\[-2pt]
        N_{\!DP-System}  &:\; \text{Number of data-parallel systems} \\[-2pt]
        Lat_{\!Pref} &:\; \text{Average prefill latency} \\[-2pt]
        Lat_{\!Dec} &:\; \text{Average decode latency}\\[-2pt]
    \end{aligned}
    \notag
    \end{align*}
}

\noindent {\bf Throughput trends with TP.} As Figures~\ref{fig:llama3_70b_tput} and~\ref{fig:llama3_405b_tput} show, the TP and hybrid TP \& PP setups degrade the TPS values compared to PP. Although TP shortens the execution of Attention and FFN blocks by sharding their compute workloads across multiple GPUs, it introduces the overhead of all-reduce operations. Even though higher TP sizes reduce the pass time through transformer blocks by involving more GPUs, the rate of latency reduction is not commensurate with the size of parallelization for the same batch size, resulting in system TPS degradation. That is, the amount of latency reduction when switching from a setup without parallelization to any size of TP is not inversely proportional to the increasing rate of TP size due to the overhead of all-reduce operations, whose contribution to TTFT values is illustrated in Figure~\ref{fig:tp_depth} for various TP sizes.

\noindent {\bf Throughput trends with PP.} Deploying Llama 3.1-70B and -405B models with the PP scheme improves both models' system throughput by increasing nano batch sizes\footnote{\scriptsize Global batch size refers to total prompts served actively by the system, while nano and micro batch sizes stand for the batches per-GPU and per-node at inference time.} up to the saturation level at which decode operations become purely compute-bound, likewise the prefills. Distributing transformers across pipeline stages in PP mitigates the burden of model weights per GPU and breaks the memory-bound behavior across decode phases due to more available space for KV cache, allowing larger nano batches, and thereby improving compute utilization. To illustrate, an inference system operating on a node without any model parallelism can batch at most 32 prompts from the MLPerf dataset per GPU, where all GPUs serve inferences through the Llama 3.1-405B model in a data-parallel manner, while PP4 and PP8 settings improve the maximum nano batch size to 256 and 512, respectively, as shown in Figure~\ref{fig:llama3_405b_tput}. 

Although PP8 supports a 16 times larger batch size, it delivers only a 1.35× TPS gain over the only data-parallel scenario (i.e., without TP or PP involvement). The relatively smaller TPS gain, as opposed to a significant increase in batch size, indicates that larger batch sizes push the decode workloads into a more compute-bound regime, which is already the case with prefill. Growing a compute-bound workload without allocating additional compute resources proportionally increases both TTFT and TPOT latencies, and thus the denominator of the TPS formula. For instance, the average prefill and decode durations in the PP8 setup are 16.03× and 4.24× higher, respectively, than their respective durations in the data-parallel-only scheme. Additionally, due to significantly larger ISL values compared to OSL for summarization tasks, as represented by the MLPerf dataset, the average prefill latency dominates the inference serving and the denominator of the TPS formula. Therefore, the TPS gain from increasing the PP depth and nano batch size while serving a dataset with much longer ISL over OSL is smaller than that of a dataset with comparably similar ISL and OSL values. As a result, Llama 3.1-70B configured with the PP8 scheme and a batch size of 1024 achieves 4.2× and 37× higher TPS than the scenario without model parallelism and a batch size of 1 when serving inferences for the LongAlpaca and combined datasets, respectively. In addition, the 405B version using PP8 and batch sizes of 512 settings delivers 13.8× and 44× TPS improvements on the MLPerf and combined datasets, respectively, in comparison with a system with no-model parallelism and a batch size of 1. 


%% file: sections/disc_and_conc.tex

\section{Discussion} \label{sec:disc}
\noindent {\bf Performance trends for more-sophisticated models.}
Given the growing popularity of Mixture-of-Experts (MoE) models~\cite{moe_survey}, extending our analysis of parallelization strategies to encompass these architectures would be highly beneficial. In MoE transformers, a gating layer positioned after the attention block selects the top-$k$ most suitable experts from a pool of fully connected networks and routes the forward pass through them. For an MoE model deployed with a TP scheme, the participant GPUs calculate the gating scores locally, and the routing decisions to activate experts for each individual token are obtained through an additional all-to-all communication. Still, the core attention block and expert layers are sharded across GPUs participating in the TP. Therefore, TP can potentially provide significant latency flexibility for MoE models, as in dense LLMs, even though the MoE transformer incurs an extra overhead of all-to-all communication. Additionally, PP, which distributes transformer blocks among GPUs, is orthogonal to updated transformers of MoE models, and thus potentially improves the system throughput, especially for models that barely fit into device memory. 

\noindent {\bf Performance trends within multi-node systems.} Determining the performance trends of model-parallelism techniques within multi-node setups will further contribute to navigating the optimal LLM-inference deployments at larger scales. \cite{nccl_multinode} probes communication performances across intra- and inter-node setups, where the intra-node has 150 GB/s all-to-all connected links across GPUs, whereas the inter-node setup relies on a DragonFly~\cite{dragonFly} topology linking nodes with the bandwidth of 25 GB/s. The study reveals that the performance of the ring all-reduce algorithm within intra-node setup is 6-8× faster. Thus, increasing the TP size to a multi-node scale is expected to significantly amplify the latency contribution of all-reduce kernels to TTFT and TPOT, causing inter-node communication to become the major latency bottleneck for LLM inference, as in training~\cite{distributedTrans}. Similarly, PP within a multi-node system will increase the latency experienced between pipeline stages, thereby negatively affecting latency objectives. Moreover, elevated inference latency, which increases the denominator of the throughput formula, slightly harms system throughput due to the elevated communication burden associated with model parallelism in a multi-node system. 

\noindent {\bf Novel Insights.}
One familiar with the parallelization deployment discussions in Section~\ref{sec:parStrategies} can infer that TP introduces latency flexibility, while PP offers a throughput increase but no latency benefit. Beyond these intuitive deductions, we observe that TP, dedicating more compute resources to a transformer pass, scales down TTFT almost linearly but shows weaker scaling for TPOT due to the compute-heavy nature of prefill relative to decode, and all-reduce becomes a significant latency bottleneck in both latency metrics for larger TP sizes, in particular at multi-node scale. We additionally reveal that PP offers limited throughput scaling as the system throughput saturates when the decode becomes fully compute-bound for deeper parallelization levels. The underlying reason is that PP maintains the compute resource devoted to a transformer pass as constant, in contrast to TP, but instead allows increasing per-GPU batch sizes. However, using larger batches within fully compute-bound inference workloads grows the TTFT and TPOT latencies proportionally when configured with PP, resulting in saturated output TPS. 

\section{Concluding Remarks} \label{sec:conc}
This paper thoroughly quantifies the LLM-powered application objectives--latency and throughput--over existing parallelization strategies, their hybrid usages and various depths, batching preferences, and sequence lengths, by configuring them in our in-house simulator with diverse inference configurations that deploy dense LLMs--Llama 3.1-70B and 405B--for the cornerstone datasets, capturing both long and short sequences. Our analysis concludes that Tensor Parallelism (TP) substantially reduces TTFT and TPOT latencies in comparison with Pipeline Parallelism (PP) and no-model-parallelism choices, with deeper TP presenting further latency flexibility. We further depict the contribution of all-reduce kernels incurred by TP to the TTFT for various TP degrees and reveal the latency behavior for changing aggregate link speeds. Moreover, our analysis shows that PP delivers higher throughput than TP across all batch sizes and surpasses the only data-parallel baseline at larger batches, primarily due to its greater KV-cache capacity, where inter-pipeline communication adds far less latency than all-reduce operations introduced in TP. Consequently, TP at different sizes offers flexibility in meeting latency objectives, whereas PP is better suited for throughput-oriented scenarios. Thus, a hybrid use of TP \& PP provides a tunable balance, allowing latency control through the TP degree and throughput scaling through the PP depth.
